%% file: bsmumu2fb_v09.tex
\documentclass[superscriptaddress,preprintnumbers,showpacs,amsmath,amssymb,prl,twocolumn]{revtex4}

\usepackage{graphicx}
\usepackage{dcolumn}
\usepackage{bm}
\usepackage{ifthen}
\usepackage{setspace}
\usepackage{color}

\def\ra{\rightarrow}
\newcommand{\GeV}{\ensuremath{\mathrm{Ge\kern -0.1em V}}}
\newcommand{\MeV}{\ensuremath{\mathrm{Me\kern -0.1em V}}}

\newcommand{\pp}{\ensuremath{p\overline{p}}}
\newcommand{\pb}{\ensuremath{\mathrm{pb}^{-1}}}
\newcommand{\fb}{\ensuremath{\mathrm{fb}^{-1}}}
\newcommand{\bs}{\ensuremath{B_s^0}}
\newcommand{\bd}{\ensuremath{B^0}}
\newcommand{\bsd}{\ensuremath{B_s^0(B^0)}}
\newcommand{\bu}{\ensuremath{B^{+}}}

\newcommand{\mm}{\ensuremath{\mu^{+}\mu^{-}}}

\newcommand{\bsmm}{\ensuremath{\bs\ra\mm}}
\newcommand{\bdmm}{\ensuremath{\bd\ra\mm}}
\newcommand{\bsdmm}{\ensuremath{\bsd\ra\mm}}
\newcommand{\bjk}{\ensuremath{\bu\ra J/\psi K^{+}}}
\newcommand{\bjpi}{\ensuremath{\bu\ra J/\psi \pi^{+}}}
\newcommand{\bsjp}{\ensuremath{\bs\ra J/\psi \phi}}

\newcommand{\brbsmm}{\ensuremath{\mathcal{B}(\bsmm)}}
\newcommand{\brbdmm}{\ensuremath{\mathcal{B}(\bdmm)}}
\newcommand{\brbsdmm}{\ensuremath{\mathcal{B}(B_s^0(B^0)\ra\mm)}}
\newcommand{\Mmm}{\ensuremath{m_{\mu\mu}}}

\newcommand{\ctau}{\ensuremath{\lambda}}
\newcommand{\pting}{\ensuremath{\Delta\Theta}}
\newcommand{\iso}{\ensuremath{\mathit{I}}}
\newcommand{\nn}{\ensuremath{\nu_{N}}}
\newcommand{\ptmm}{\ensuremath{\vec{p}^{\:\mu\mu}_{T}}}
\newcommand{\pmm}{\ensuremath{\vec{p}^{\:\mu\mu}}}
\newcommand{\cdf}{CDF~II}

\newcommand{\dedx}{\ensuremath{dE/dx}}

\begin{document}
% ============================================================================
% --- TITLE PAGE
% ============================================================================

%{\flushright $~$\\Version 08a Nov, 30 2007 $~~~~~~~~~$ CDF NOTE 9013\\
%}

\title{

\bf{ Search for $\bsmm$ and $\bdmm$ Decays 
        with  $2\:\fb$ of ${p\overline{p}}$ Collisions}}

%\collaboration{CDF Collaboration}
%\noaffiliation

\begin{abstract}
  We have performed a search for \bsmm\ and \bdmm\ decays in \pp\ collisions
  at $\sqrt{s} = 1.96$~TeV using $2\:\fb$ of integrated luminosity collected by the \cdf\ 
  detector at the Fermilab Tevatron Collider.
  The observed number of 
  \bs\ and \bd\ candidates is consistent with background expectations.   The 
  resulting upper limits on the branching fractions are 
  $\brbsmm < 5.8\times10^{-8}$ and $\brbdmm < 1.8\times10^{-8}$ at 
  $95\%$ C.L.
\end{abstract}

\input{July_2007_Authors_Visitors}


\maketitle

% ============================================================================
% --- BODY OF PAPER
% ============================================================================

Processes involving flavor-changing neutral currents (FCNCs) provide excellent
signatures with which to search for evidence of new physics since they have small branching fractions in the
standard model (SM), while new physics contributions can provide a significant enhancement.
The FCNC decays $\bs (\bd) \rightarrow \mu^+\mu^-$~\cite{qconj}
occur in the SM only through higher order diagrams.
% and are
%further suppressed by the helicity factor, $(m_\mu/m_B)^2$. 
%The $\bd$ decay is also suppressed with respect to the $\bs$ decay
%by the ratio of CKM elements, $\left|V_{td}/V_{ts}\right|^2$.
The SM expectations for these branching fractions are 
$\brbsmm = (3.42\pm0.54)\times10^{-9}$ and $\brbdmm = (1.00\pm0.14)\times10^{-10}$~\cite{smbr}, 
which are one order of magnitude smaller
than current experimental sensitivity. Previous  
bounds, based on $1.3\:\fb$ and $364\:\pb$ are $\brbsmm <1.2\times10^{-7}$ and
$\brbdmm <5.1\times 10^{-8}$ at the 95\% C.L., respectively~\cite{D0prdrc,PRL2}.

Enhancements to \bsdmm\ occur in many new-physics models.
In supersymmetry (SUSY) models, contributions from diagrams including
supersymmetric particles can increase \brbsdmm\ by 
several orders of magnitude at large $\tan\beta$, the ratio of 
vacuum expectation values of the Higgs doublets~\cite{tanb}.
In the minimal supersymmetric standard model (MSSM), the enhancement is proportional to
$\tan^6\!\beta$.  Global analyses
including all existing experimental constraints suggest that the large $\tan\!\beta$ 
region is of interest~\cite{rparity,dark,chi2}.
In contrast,
SUSY R-parity violating models~\cite{rparity} and non-minimal flavor violating models~\cite{nmfv}
can both enhance \bsmm\ and \bdmm\ separately even at low $\tan\!\beta$.
In the absence of an observation, limits on \brbsmm\ are complementary to those 
provided by other experimental measurements, and together would significantly constrain 
the allowed supersymmetric parameter space.
For example, if the lightest neutralino in SUSY models is a
cold dark matter (CDM) particle, \brbsmm\ and
constraints on the amount of CDM in the universe from cosmic microwave anisotropy measurements
can be exploited in this way~\cite{rparity,dark,chi2}.
Then, for instance, in minimal supergravity (mSUGRA)
models limits on \brbsmm\ will correspond to bounds on
superpartner particle masses that are beyond the sensitivity of the corresponding
direct searches for those particles in colliding-beam experiments~\cite{rparity}.
In general, the search for these rare
decays is central to exploring a large class of new-physics models.

This measurement uses
$2\:\fb$ of integrated luminosity collected by the upgraded Collider Detector at Fermilab (\cdf) and
supersedes our previous measurement using 
$364\:\pb$~\cite{PRL2}. 
The sensitivity of the analysis is improved significantly by increasing the
integrated luminosity of the event sample, using an enhanced muon selection, employing a
neural network (NN) discriminant to separate signal from background, and performing the search in a
two dimensional grid in dimuon mass and NN space.

A detailed description of the \cdf\ detector can be found in Ref.~\cite{Acosta:2004yw}. 
%Here we give a brief description of the elements used in
%this analysis.
Charged particle tracking is provided by a silicon microstrip detector 
surrounded by an open-cell wire drift chamber immersed in a 1.4 T solenoidal 
magnetic field. This system provides precise vertex determination and momentum 
measurements for charged particles in a pseudorapidity range 
$\left|\eta\right|<1.0$, where $\eta = -\ln(\tan\frac{\theta}{2})$ and 
$\theta$ is the angle of the track measured with respect to the proton beam direction.
Additionally, the drift chamber measures the ionization per unit path length, $\dedx$,
for particle identification.  
Surrounding the
tracking detectors are electromagnetic and hadronic sampling calorimeters
arranged in a projective geometry.
Drift chambers referred to as CMU and CMX are
located behind the calorimeters to detect muons
within
$\left|\eta\right|<0.6$ and $0.6<\left|\eta\right|<1.0$, respectively.  

Events are recorded for subsequent analysis if they
have either of two topologies that satisfy the requirements
of the online trigger system:
CMU-CMU, which includes events where both muon candidates are triggered using
the central muon detectors, and CMU-CMX,
which includes events where one of the muons is triggered
in the central muon detector and one in the higher pseudorapidity region.
%Since the expected
%signal-to-background ratios are different between the two trigger classes, they are
%treated and optimized separately throughout the analysis, and the limits are combined at the end.
The details of the trigger 
system and selection requirements can be found in Refs.~\cite{PRL1,Acosta:2004yw}.  

In the offline analysis, the trigger selection is refined by applying a series of
``baseline'' requirements.
%that substantially reduce the backgrounds while
%preserving the majority of the signal.
We select two oppositely charged 
muon candidates within a dimuon invariant mass window of $4.669 < \Mmm < 5.969\:\GeV/c^2$ around
the $\bs$ and $\bd$ masses. 
Backgrounds from hadrons misidentified as muons
are suppressed by selecting muon candidates using a likelihood
function.  This function tests the consistency of electromagnetic and
hadronic energy with that expected
for a minimum 
ionizing particle
and the differences between extrapolated track trajectories and muon system hits~\cite{mixprl}.
In addition, backgrounds from kaons that penetrate through the calorimeter to the muon system or decay in flight outside
the drift chamber are further suppressed by
a loose selection based on $\dedx$~\cite{bhhprl}.
The inputs to the muon likelihood and the $\dedx$ performance are calibrated using
samples of $J/\psi\ra\mm$, $D^0 \rightarrow K^- \pi^+$ and $\Lambda \rightarrow p\pi^-$ decays.
To reduce combinatorial backgrounds the muon candidates are required to have transverse momentum 
relative to the beam direction $p_{T}>2.0(2.2)\;\GeV/c$ for CMU(CMX),
and  $| \ptmm | > 4\;\GeV/c$, where $\ptmm$ is the transverse component of the
sum of the muon momentum vectors.
The remaining pairs of muon tracks are fit under the constraint that they
come from the same three-dimensional (3D) space point.
To achieve further separation of signal from background, 
we employ additional discriminating variables.
As in our previous analysis~\cite{PRL2} these include 
the measured proper decay time, $\ctau$; the proper decay time divided by the estimated uncertainty, 
$\ctau / \sigma_{\ctau}$; the 3D opening angle 
between vectors \pmm\ and the displacement vector between the primary vertex and the dimuon vertex, $\pting$;
and the {\it B}-candidate track isolation, $\iso$~\cite{isodef}.  We require 
that $\ctau / \sigma_{\ctau} > 2$,
$\pting < 0.7~\rm{rad}$, and $\iso > 0.50$.
There are 30 666 dimuon candidates 
that fulfill the above trigger and baseline reconstruction requirements.  
The baseline selection reduces combinatorial backgrounds 
by a factor of 300 while keeping approximately 50\% 
of the signal events that are within the acceptance (geometric and kinematic requirements) of the trigger.
Relative to the previous
analysis, the application of the muon likelihood and $\dedx$ selection is approximately 97\% efficient 
for the signal and reduces
combinatorial backgrounds by 35\% and constrains backgrounds from two-body
hadronic $B$ decays to a level that has negligible impact on the analysis.
A sample of $\bjk$ events is collected to serve as a normalization mode using
the same baseline
requirements, but including a requirement of $p_T>1~\GeV/c$ for the kaon  
candidate and constructing the $\bjk\ra\mm$ vertex using only the muon candidate tracks.

\begin{figure}[top]
  \centering
  \vspace*{-0.0in}
  \includegraphics[width=3.2in]{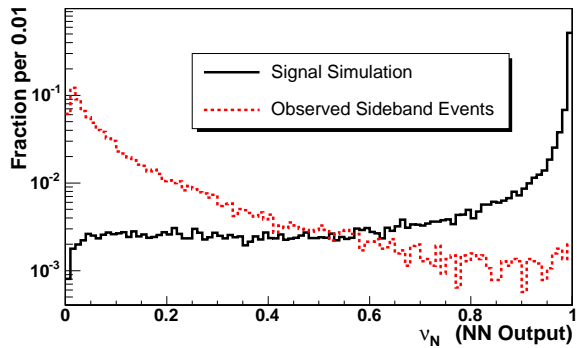}
  \caption{\label{fig:sigbgNN} Distributions of \nn\ for simulated \bsmm\ signal and observed
sideband events.} 
\end{figure}

For the final event selection we use the following discriminating variables:
\Mmm , \ctau , $\ctau / \sigma_{\ctau}$, \pting , \iso , $| \ptmm |$, 
and the $p_T$ of the lower momentum muon
candidate.  
To enhance signal and background separation we construct a
NN discriminant, \nn, based on all the discriminating variables except $\Mmm$, which
is used to define signal and sideband background regions.
The NN is trained using background events sampled from the sideband regions
and signal events generated with a simulation described below.
The \nn\ distributions of $\bs$ signal and sideband background
events are shown in Fig.~\ref{fig:sigbgNN}.  
Only part of the total number of background and simulated signal
events are used in order to have unbiased samples to test
the background discrimination and signal efficiency.  
%We also conducted tests
%for biases as a function of mass, which could enhance a false signal, and none were found.

For the final selection, we define search regions of
$5.310  < \Mmm < 5.430\:\GeV/c^2$ for the $\bs$ and $5.219  < \Mmm < 5.339\:\GeV/c^2$ for the $\bd$ around the
mass values  $m_{\bs} = 5.370\:\GeV/c^2$ and $m_{\bd} = 5.279\:\GeV/c^2$~\cite{PDG2006}, respectively. These regions correspond 
to approximately $\pm 2.5$ times $\sigma_{m}$, the estimated two-track invariant mass resolution, 
where $\sigma_{m}\approx24$~$\mathrm{MeV}/c^{2}$.
The sideband regions $4.669 < \Mmm < 5.169\:\GeV/c^2$ and 
$5.469 < \Mmm < 5.969\:\GeV/c^2$ are used to estimate the combinatorial backgrounds in the
signal regions.  Backgrounds from the two-body hadronic $\bs$ and $\bd$ decays, 
$B \ra h^+h^-$, where $h^{\pm}$ are $\pi^{\pm}$ or $K^{\pm}$, 
which peak in the $\bs$ and $\bd$ invariant mass signal region
and do not occur in the sidebands, are estimated separately.
The content of signal regions were not
unveiled until all selection criteria were finalized.
The final selection criteria were determined from an {\it a priori}
optimization, which maximizes sensitivity of the expected limit.  
The kinematics of $\bsmm$ and $\bdmm$ decays
are similar enough that the efficiencies are the same within statistical uncertainties and
the same efficiencies and optimizations are used.

For measuring efficiencies, estimating backgrounds, and optimizing the
analysis, samples of $\bsdmm$, $\bjk$, and $B \ra h^+h^-$ 
are generated with the {\sc pythia} simulation 
program~\cite{pythia} and a \cdf\ detector simulation.
The $B$-hadron $p_{T}$ 
spectrum and the \iso\ distribution of the $B$-hadrons 
are weighted to match distributions measured in samples of
\bjk\ and \bsjp\ events~\cite{Acosta:2004yw,mixprl}.

We use a relative normalization to determine the \bsmm\ branching
fraction:
\begin{eqnarray}\label{eq:brbsmm}
\brbsmm = \frac{N_s}{N_{+}}\cdot
     \frac{\alpha_{+}}{\alpha_{s}}
     \cdot\frac{\epsilon_{+}}{\epsilon_{s}} \cdot 
     \frac{1}{\epsilon_{N}} \cdot \nonumber
     \frac{f_{u}}{f_{s}}  
     \cdot \mathcal{B}(B^+),~~~ \rm{(1)}
\end{eqnarray}
where  $N_s$ is the number of \bsmm\ candidate events.
We observe $N_{+} = 11~387\pm164$ \bjk\ candidates.
This number has been corrected for background using sideband subtraction;
the contribution of \bjpi\ events is negligible.
We use $\mathcal{B}(B^+) = \mathcal{B}(\bjk\ra\mm K^{+})=(5.94\pm0.21)\times 10^{-5}$~\cite{PDG2006} and
the ratio of $B$-hadron production fractions ${f_{u}}/{f_{s}}=3.86\pm0.59$~\cite{PDG2006}. 
The parameter $\alpha_{s}$ ($\alpha_{+}$) is the acceptance of the trigger and  
$\epsilon_{s}$ ($\epsilon_{+}$) is the efficiency of the
reconstruction requirements for the signal (normalization) mode.  The reconstruction efficiency
includes trigger, track, muon, and baseline selection efficiencies.  
The NN efficiency, ${\epsilon_{N}}$, only applies to the signal mode.
The expression for \brbdmm\ is derived by replacing \bs\ with \bd\ 
and the fragmentation ratio with ${f_{u}}/{f_{d}}=1$.

The ratios of acceptances ${\alpha_{+}}/{\alpha_{s}}$ are 
$0.297\pm0.020$ and $0.191\pm0.013$ for the CMU-CMU and CMU-CMX trigger 
channels, respectively.  These ratios are measured using simulated events, and 
the uncertainties include contributions from systematic variations
of the modeling of the $B$-hadron $p_{T}$ distributions, the longitudinal 
beam profile, and from the statistics of the simulated event samples. 
The ratio of reconstruction efficiencies is ${\epsilon_{+}}/{\epsilon_{s}}=0.89\pm0.04$.
Muon reconstruction efficiencies are estimated as a function of muon $p_T$  
using observed event samples of inclusive $J/\psi\ra\mm$ decays.
Systematic uncertainties in the efficiency ratio, 
${\epsilon_{+}}/{\epsilon_{s}}$, largely 
cancel with the exception of the kaon efficiency from the \bu\ decay.
The uncertainty 
is dominated by kinematic differences
between inclusive $J/\psi\ra\mm$ and \bsdmm\ decays.
The efficiency, $\epsilon_{N}$, is estimated from the simulation.  
We assign a relative systematic uncertainty on $\epsilon_{N}$ of 6\%\ 
based on comparisons of NN performance in simulated and observed $\bjk$ event samples and 
the statistical uncertainty on studies of the $\bs$ $p_T$ and $\iso$ distributions from observed $\bsjp$ event samples.  
The NN provides approximately 25\% better background rejection for the same signal efficiency
compared to the previous analysis~\cite{PRL2}.

The expected background is obtained by summing contributions from
the combinatorial continuum and from $B\rightarrow h^+h^-$ decays.
The contribution from
other heavy-flavor decays is negligible.  We estimate the combinatorial
background by linearly extrapolating from the sideband region to the
signal region.
The $B\rightarrow h^+h^-$ contributions are about a factor of ten smaller
than the combinatorial background and are estimated using efficiencies
taken from the simulation,  probabilities of misidentifying hadrons as muons measured in
a $D^0 \rightarrow \pi K$ data sample, and normalizations derived from branching 
fractions from Refs.~\cite{bhhprl,PDG2006}.
The two-body invariant mass distribution of the simulated $B\rightarrow h^+h^-$ candidates is
calculated from the momentum of the hadrons assuming the muon mass
hypothesis.
The background estimates are cross-checked using
three independent control samples: $\mu^{\pm}\mu^{\pm}$ events, \mm\ events with $\ctau<0$, and
a misidentified muon-enhanced \mm\ sample in which we require one muon 
candidate to fail the muon quality requirements.  
We compare the predicted and observed
number of events in these samples for a wide range of \nn\
requirements and observe no significant discrepancies.

\begin{figure}[]
  \centering
  \includegraphics[width=3.2in]{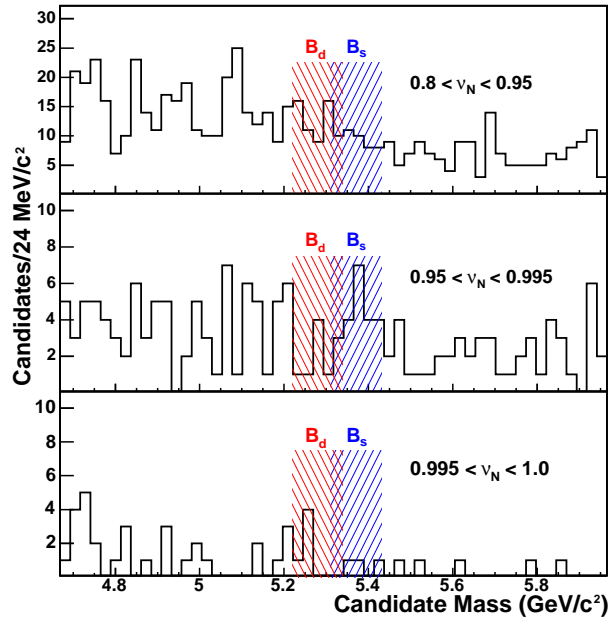}
  \caption{\label{fig:MvL80} The \mm\ invariant mass distribution for events 
   satisfying all selection criteria for the final three ranges of $\nu_{N}$.}
\end{figure}

Using an {\it a priori} optimization procedure, we found that subdividing the signal
region into several \nn\ and mass bins to exploit the shape of the mass distribution
and the higher signal to background ratios for the higher \nn\ values 
improves the sensitivity by 15\% relative
to using a single bin.
The signal region is divided into five equal mass bins of  $24$~$\mathrm{MeV}/c^{2}$ and
three \nn\ bins delineated at $0.8$, $0.95$, $0.995$ and $1.0$.  
The backgrounds, efficiencies, and limits are calculated in each bin
separately.  
Summing over the mass bins in each slice of $\nu_{N}$,
the corresponding ${\epsilon_{N}}$s are estimated to be
12\%, 23\%, and 44\% and the expected SM yields of \bsmm\ events are $0.08\pm0.03$, $0.15\pm0.05$, and $0.30\pm0.10$, respectively.
The expected yield of \bdmm\ events is ten times smaller.
Using these optimized selection criteria, we
compute an expected limit of $\brbsmm < 4.9 \times 10^{-8}$ at 95\% C.L.
The expected limit is a factor of five better than the expected limit of the previous
analysis~\cite{PRL2}.
The limits are estimated from Eq.~(\ref{eq:brbsmm}) using
the confidence level (CLs) method of Ref.~\cite{PDG2006} to extract the 95\% C.L. upper bound on $N_s$;
the limit
incorporates Gaussian uncertainties on the signal acceptance and efficiency
as well as the background estimates.
The number of observed events is compared with the total predicted
background in Table~\ref{tab:resultsUU} for each bin of mass and  $\nu_{N}$.
The uncertainty on the background estimate is dominated by the statistical uncertainty
of the sideband sample. 
The $\mm$ invariant mass distributions for the three different $\nu_{N}$ ranges
are shown in Fig.~\ref{fig:MvL80}. 
The observed event rates are
consistent with SM background expectations.
We extract 95\% (90\%) C.L. limits of
$\brbsmm < 5.8\times 10^{-8}$ $(4.7\times 10^{-8})$ and
$\brbdmm < 1.8\times 10^{-8}$ $(1.5\times 10^{-8})$. 

% --- Box Open: 2D Mass vs LH for LH>0.90
%%BoundingBox:  40 20 547 547

\begin{table}[top]
  \begin{center} 
%\scriptsize
    \begin{tabular}{rccccc}
      \hline
      \hline
     
      \bs\ ~~ & \multicolumn{5}{c}{mass bins} \\
      \nn\  ~  & I & II & III & IV & V \\
      \hline
      A Exp     & 10.3$\pm$0.5 & 10.1$\pm$0.4 & 9.9$\pm$0.4 &9.7$\pm$0.4 & 9.5$\pm$0.4 \\
      Obs      & 11& 9& 10& 9& 5\\
      B Exp    & 3.7$\pm$0.3& 3.7$\pm$0.3& 3.6$\pm$0.3 &3.5$\pm$0.3 & 3.5$\pm$0.3\\
      Obs      & 4& 3& 6& 6& 2 \\
      C Exp     & 0.7$\pm$0.1& 0.7$\pm$0.1& 0.7$\pm$0.1& 0.7$\pm$0.1& 0.7$\pm$0.1\\
      Obs      & 0& 1& 1& 0& 1\\
\\
      \bd\ ~~ &  &  &  &  &  \\
      \hline
      A Exp& 11.0$\pm$0.6 & 10.8$\pm$0.5 & 10.7$\pm$0.5 &10.5$\pm$0.5 & 10.3$\pm$0.5 \\
        Obs & 15& 13& 9& 14& 9\\
      B Exp & 4.0$\pm$0.3& 3.9$\pm$0.3 &3.9$\pm$0.3 &3.8$\pm$0.3 & 3.7$\pm$0.3\\
      Obs & 1& 1& 5& 2& 4 \\
      C Exp & 0.8$\pm$0.1& 0.8$\pm$0.1& 0.8$\pm$0.1& 0.8$\pm$0.1& 0.8$\pm$0.1\\
      Obs     & 2& 3& 1& 0& 0\\
      \hline
      \hline

    \end{tabular}

    \caption{\label{tab:resultsUU} The total number of expected(Exp) and observed(Obs)
     background events for the \bs\ (upper) 
     and \bd\ (lower) signal windows.   
     \nn\ bins:  
     A (0.80-0.95), B (0.95-0.995), C (0.995-1.0)  and five equal sized mass bins, (I-V),
     as described in the text.
    }
  \end{center}
\end{table}

In mSUGRA, branching ratios as low as 
$\brbsmm = 5.8\times 10^{-8}$ occur for common gaugino mass 
parameter, $m_{1/2}$, below $380\:\GeV/c^2$ at $\tan\!\beta = 50$ in
a CDM-allowed co-annihilation region~\cite{mSUGRA}.  In this scenario we exclude 
gluino masses below $925\:\GeV/c^2$.

%This $m_{1/2}$
%limit corresponds to a limit on the supersymmetric gluon mass of $925\:\GeV/c^2$, which is beyond the kinematic reach of
%the Tevatron.
This Letter reports a search for the rare FCNC decays \bsmm\ and \bdmm\
with $2\:\fb$ integrated luminosity collected in $p\overline{p}$ collisions at $\sqrt{s} = 1.96$~TeV using the
\cdf\ detector and employing improved analysis techniques.
We observe no evidence for
new physics and set limits that are
the most stringent
to date, improving the previous results~\cite{D0prdrc,PRL2} by a factor of two or more.
These limits place further constraints on
new-physics models~\cite{tanb,rparity,dark,chi2,nmfv}, and complement direct searches for new physics.
We expect the analysis sensitivity to continue to improve as we include larger
data sets.

We thank the Fermilab staff and the technical staffs of the participating institutions for their vital contributions. This work was supported by the U.S. Department of Energy and National Science Foundation; the Italian Istituto Nazionale di Fisica Nucleare; the Ministry of Education, Culture, Sports, Science and Technology of Japan; the Natural Sciences and Engineering Research Council of Canada; the National Science Council of the Republic of China; the Swiss National Science Foundation; the A.P. Sloan Foundation; the Bundesministerium f\"ur Bildung und Forschung, Germany; the Korean Science and Engineering Foundation and the Korean Research Foundation; the Science and Technology Facilities Council and the Royal Society, UK; the Institut National de Physique Nucleaire et Physique des Particules/CNRS; the Russian Foundation for Basic Research; the Comisi\'on Interministerial de Ciencia y Tecnolog\'{\i}a, Spain; the European Community's Human Potential Programme; the Slovak R\&D Agency; and the Academy of Finland. 

% ============================================================================
% --- BIBLIOGRAPHY
% ============================================================================

%%%\bibliography{prl}

% ============================================================================
\end{document}

%% file: July_2007_Authors_Visitors.tex
\affiliation{Institute of Physics, Academia Sinica, Taipei, Taiwan 11529, Republic of China} 
\affiliation{Argonne National Laboratory, Argonne, Illinois 60439} 
\affiliation{Institut de Fisica d'Altes Energies, Universitat Autonoma de Barcelona, E-08193, Bellaterra (Barcelona), Spain} 
\affiliation{Baylor University, Waco, Texas  76798} 
\affiliation{Istituto Nazionale di Fisica Nucleare, University of Bologna, I-40127 Bologna, Italy} 
\affiliation{Brandeis University, Waltham, Massachusetts 02254} 
\affiliation{University of California, Davis, Davis, California  95616} 
\affiliation{University of California, Los Angeles, Los Angeles, California  90024} 
\affiliation{University of California, San Diego, La Jolla, California  92093} 
\affiliation{University of California, Santa Barbara, Santa Barbara, California 93106} 
\affiliation{Instituto de Fisica de Cantabria, CSIC-University of Cantabria, 39005 Santander, Spain} 
\affiliation{Carnegie Mellon University, Pittsburgh, PA  15213} 
\affiliation{Enrico Fermi Institute, University of Chicago, Chicago, Illinois 60637} 
\affiliation{Comenius University, 842 48 Bratislava, Slovakia; Institute of Experimental Physics, 040 01 Kosice, Slovakia} 
\affiliation{Joint Institute for Nuclear Research, RU-141980 Dubna, Russia} 
\affiliation{Duke University, Durham, North Carolina  27708} 
\affiliation{Fermi National Accelerator Laboratory, Batavia, Illinois 60510} 
\affiliation{University of Florida, Gainesville, Florida  32611} 
\affiliation{Laboratori Nazionali di Frascati, Istituto Nazionale di Fisica Nucleare, I-00044 Frascati, Italy} 
\affiliation{University of Geneva, CH-1211 Geneva 4, Switzerland} 
\affiliation{Glasgow University, Glasgow G12 8QQ, United Kingdom} 
\affiliation{Harvard University, Cambridge, Massachusetts 02138} 
\affiliation{Division of High Energy Physics, Department of Physics, University of Helsinki and Helsinki Institute of Physics, FIN-00014, Helsinki, Finland} 
\affiliation{University of Illinois, Urbana, Illinois 61801} 
\affiliation{The Johns Hopkins University, Baltimore, Maryland 21218} 
\affiliation{Institut f\"{u}r Experimentelle Kernphysik, Universit\"{a}t Karlsruhe, 76128 Karlsruhe, Germany} 
\affiliation{Center for High Energy Physics: Kyungpook National University, Daegu 702-701, Korea; Seoul National University, Seoul 151-742, Korea; Sungkyunkwan University, Suwon 440-746, Korea; Korea Institute of Science and Technology Information, Daejeon, 305-806, Korea; Chonnam National University, Gwangju, 500-757, Korea} 
\affiliation{Ernest Orlando Lawrence Berkeley National Laboratory, Berkeley, California 94720} 
\affiliation{University of Liverpool, Liverpool L69 7ZE, United Kingdom} 
\affiliation{University College London, London WC1E 6BT, United Kingdom} 
\affiliation{Centro de Investigaciones Energeticas Medioambientales y Tecnologicas, E-28040 Madrid, Spain} 
\affiliation{Massachusetts Institute of Technology, Cambridge, Massachusetts  02139} 
\affiliation{Institute of Particle Physics: McGill University, Montr\'{e}al, Canada H3A~2T8; and University of Toronto, Toronto, Canada M5S~1A7} 
\affiliation{University of Michigan, Ann Arbor, Michigan 48109} 
\affiliation{Michigan State University, East Lansing, Michigan  48824} 
\affiliation{University of New Mexico, Albuquerque, New Mexico 87131} 
\affiliation{Northwestern University, Evanston, Illinois  60208} 
\affiliation{The Ohio State University, Columbus, Ohio  43210} 
\affiliation{Okayama University, Okayama 700-8530, Japan} 
\affiliation{Osaka City University, Osaka 588, Japan} 
\affiliation{University of Oxford, Oxford OX1 3RH, United Kingdom} 
\affiliation{University of Padova, Istituto Nazionale di Fisica Nucleare, Sezione di Padova-Trento, I-35131 Padova, Italy} 
\affiliation{LPNHE, Universite Pierre et Marie Curie/IN2P3-CNRS, UMR7585, Paris, F-75252 France} 
\affiliation{University of Pennsylvania, Philadelphia, Pennsylvania 19104} 
\affiliation{Istituto Nazionale di Fisica Nucleare Pisa, Universities of Pisa, Siena and Scuola Normale Superiore, I-56127 Pisa, Italy} 
\affiliation{University of Pittsburgh, Pittsburgh, Pennsylvania 15260} 
\affiliation{Purdue University, West Lafayette, Indiana 47907} 
\affiliation{University of Rochester, Rochester, New York 14627} 
\affiliation{The Rockefeller University, New York, New York 10021} 
\affiliation{Istituto Nazionale di Fisica Nucleare, Sezione di Roma 1, University of Rome ``La Sapienza," I-00185 Roma, Italy} 
\affiliation{Rutgers University, Piscataway, New Jersey 08855} 
\affiliation{Texas A\&M University, College Station, Texas 77843} 
\affiliation{Istituto Nazionale di Fisica Nucleare, University of Trieste/\ Udine, Italy} 
\affiliation{University of Tsukuba, Tsukuba, Ibaraki 305, Japan} 
\affiliation{Tufts University, Medford, Massachusetts 02155} 
\affiliation{Waseda University, Tokyo 169, Japan} 
\affiliation{Wayne State University, Detroit, Michigan  48201} 
\affiliation{University of Wisconsin, Madison, Wisconsin 53706} 
\affiliation{Yale University, New Haven, Connecticut 06520} 
\author{T.~Aaltonen}
\affiliation{Division of High Energy Physics, Department of Physics, University of Helsinki and Helsinki Institute of Physics, FIN-00014, Helsinki, Finland}
\author{J.~Adelman}
\affiliation{Enrico Fermi Institute, University of Chicago, Chicago, Illinois 60637}
\author{T.~Akimoto}
\affiliation{University of Tsukuba, Tsukuba, Ibaraki 305, Japan}
\author{M.G.~Albrow}
\affiliation{Fermi National Accelerator Laboratory, Batavia, Illinois 60510}
\author{B.~\'{A}lvarez~Gonz\'{a}lez}
\affiliation{Instituto de Fisica de Cantabria, CSIC-University of Cantabria, 39005 Santander, Spain}
\author{S.~Amerio}
\affiliation{University of Padova, Istituto Nazionale di Fisica Nucleare, Sezione di Padova-Trento, I-35131 Padova, Italy}
\author{D.~Amidei}
\affiliation{University of Michigan, Ann Arbor, Michigan 48109}
\author{A.~Anastassov}
\affiliation{Rutgers University, Piscataway, New Jersey 08855}
\author{A.~Annovi}
\affiliation{Laboratori Nazionali di Frascati, Istituto Nazionale di Fisica Nucleare, I-00044 Frascati, Italy}
\author{J.~Antos}
\affiliation{Comenius University, 842 48 Bratislava, Slovakia; Institute of Experimental Physics, 040 01 Kosice, Slovakia}
% ...
\collaboration{CDF Collaboration\footnote{With visitors from $^a$University of Athens, 15784 Athens, Greece, 
$^b$Chinese Academy of Sciences, Beijing 100864, China, 
$^c$University of Bristol, Bristol BS8 1TL, United Kingdom, 
$^d$University Libre de Bruxelles, B-1050 Brussels, Belgium, 
$^e$University of California Irvine, Irvine, CA  92697, 
$^f$University of California Santa Cruz, Santa Cruz, CA  95064, 
$^g$Cornell University, Ithaca, NY  14853, 
$^h$University of Cyprus, Nicosia CY-1678, Cyprus, 
$^i$University College Dublin, Dublin 4, Ireland, 
$^j$University of Edinburgh, Edinburgh EH9 3JZ, United Kingdom, 
$^k$University of Heidelberg, D-69120 Heidelberg, Germany, 
$^l$Universidad Iberoamericana, Mexico D.F., Mexico, 
$^m$University of Manchester, Manchester M13 9PL, England, 
$^n$Nagasaki Institute of Applied Science, Nagasaki, Japan, 
$^o$University de Oviedo, E-33007 Oviedo, Spain, 
$^p$Queen Mary, University of London, London, E1 4NS, England, 
$^q$Texas Tech University, Lubbock, TX  79409, 
$^r$IFIC(CSIC-Universitat de Valencia), 46071 Valencia, Spain, 
}}
\noaffiliation

%% file: bsmumu2fb_v09.bbl
\begin{thebibliography}{99.}
\bibitem{qconj}
 Throughout this paper inclusion of charge conjugate reactions are implied.
\bibitem{isodef} $\iso=| \ptmm | /(\sum_{i}p^{i}_{T} + | \ptmm | )$; the
 sum is over all tracks with $\sqrt{\Delta\eta^2+\Delta\phi^2} \le 1$;
 $\Delta\phi$ and $\Delta\eta$ are the azimuthal angle and
 pseudorapidity of track $i$ with respect to \pmm.
\bibitem{mSUGRA}
 In the mSUGRA model, we assume a common trilinear parameter $A_0 = 0$ and a positive Higgsino mixing
 parameter.  In the co-annihilation region, the difference of the neutralino and stau
 masses is characterized to be less than $20\:\GeV/c^2$.  See Ref.~\cite{rparity} for details.
\end{thebibliography}
